\documentclass[11pt]{article}
\usepackage{jheppub}
\usepackage{amsmath,amssymb,amsfonts,graphicx,slashed,color}
\usepackage{subcaption}
\usepackage{tikz}
\usepackage{tikz-3dplot}
\usepackage{pgfplots}
\usepgfplotslibrary{colormaps}
\usetikzlibrary{decorations.markings}
\usetikzlibrary{decorations.pathmorphing}
\usepackage[utf8]{inputenc}
\usepackage{float}
\usepgfplotslibrary{fillbetween}
\usetikzlibrary{patterns}
\usetikzlibrary{shapes.misc}
\usepackage[utf8]{inputenc}
\usetikzlibrary{decorations.markings}
\usetikzlibrary{decorations.pathmorphing}
\newlength{\lx}
\newlength{\ly}
\newcommand{\be}{\begin{equation}}
\newcommand{\ee}{\end{equation}}
\newcommand{\bea}{\begin{eqnarray}}
\newcommand{\eea}{\end{eqnarray}}
\newcommand{\beq}{\begin{equation}}
\newcommand{\eeq}{\end{equation}}
\newcommand{\beqn}{\begin{eqnarray}}
\newcommand{\eeqn}{\end{eqnarray}}

\usepackage{upgreek}

\tikzset{->-/.style={decoration={
  markings,
  mark=at position #1 with {\arrow{>}}},postaction={decorate}}}

\tikzset{
    mark position/.style args={#1(#2)}{
        postaction={
            decorate,
            decoration={
                markings,
                mark=at position #1 with \coordinate (#2);
            }
        }
    }
}

\tikzset{
  pics/carc/.style args={#1:#2:#3}{
    code={
      \draw[pic actions] (#1:#3) arc(#1:#2:#3);
    }
  }
}

\tikzset{point/.style={insert path={ node[scale=2.5*sqrt(\pgflinewidth)]{.} }}}

\tikzset{->-/.style={decoration={
  markings,
  mark=at position #1 with {\arrow{>}}},postaction={decorate}}}

  \tikzset{-dot-/.style={decoration={
  markings,
  mark=at position #1 with {\fill[black] circle [radius=3pt,red];}},postaction={decorate}}} 
  
    \tikzset{-dotRed-/.style={decoration={
  markings,
  mark=at position #1 with {\fill[red] circle [radius=2pt,red];}},postaction={decorate}}} 
  
      \tikzset{-dotBlue-/.style={decoration={
  markings,
  mark=at position #1 with {\fill[blue] circle [radius=2pt,red];}},postaction={decorate}}} 
  
    \tikzset{-dotB-/.style={decoration={
  markings,
  mark=at position #1 with {\fill[black] circle [radius=3pt,red];}},postaction={decorate}}} 
  
    \tikzset{-dotW-/.style={decoration={
  markings,
  mark=at position #1 with {\fill[white] circle [radius=2pt,red];}},postaction={decorate}}} 

 \tikzset{-dot2-/.style={decoration={
  markings,
  mark=at position #1 with {\fill[blue] circle [radius=3pt,blue];}},postaction={decorate}}} 

    \definecolor{darkgreen}{RGB}{0,180,0}

    \definecolor{purple2}{RGB}{222,0,255}

 \tikzset{-dot3-/.style={decoration={
  markings,
  mark=at position #1 with {\fill[purple2] circle [radius=3pt,purple2];}},postaction={decorate}}} 

\tikzset{snake it/.style={decorate, decoration=snake}}

    \tikzset{cross/.style={cross out, draw=black, minimum size=2*(#1-\pgflinewidth), inner sep=0pt, outer sep=0pt},
cross/.default={3pt}}

\title{Towards an IR finite $\mathcal{S}$-matrix in the flat limit of AdS/CFT}
\author[a]{Sarthak Duary,}
\author[b]{Eliot Hijano}
\author[c]{and Milan Patra}
\affiliation[a]{International Centre for Theoretical Sciences-TIFR,
Shivakote, Hesaraghatta Hobli, Bengaluru North 560089, India}
\affiliation[b]{Department of Physics, Princeton University, Princeton, NJ 08544, USA}
\affiliation[c]{School of Physical Sciences, National Institute of Science Education and Research\\
An OCC of Homi Bhabha National Institute, Jatani-752050, India}
\emailAdd{sarthak.duary@icts.res.in}
\emailAdd{eliothijanoc@gmail.com}
\emailAdd{milan.patra@niser.ac.in}
\date{}

\abstract{ Asymptotic Fock spaces lead to IR divergences in S-matrices. The issue can be traced back to the assumption of asymptotic decoupling, and its relaxation leads to Faddeev-Kulish states and an IR-finite S-matrix. In this paper we initiate the exploration of these states in the context of the flat limit of AdS/CFT. We construct asymptotic states as flat limits of fields that interact with the electro-magnetic field in AdS, and provide a reconstruction in terms of the dual CFT operators.}

\begin{document}
\maketitle

\section{Introduction}

The appearance of IR divergences in flat spacetime is one of the features of the IR structure of QED. Physically, this is due to the presence of long-range interaction and was first studied by Weinberg \cite{Weinberg:1965nx}. The cancellation of the IR divergences was studied by restricting to sufficiently inclusive cross sections \cite{Donoghue:1999qh}. But, at the level of $\mathcal{S}$-matrix element, the IR divergence is related to the assumption of asymptotic decoupling which casts the asymptotic Hamiltonian as free. This means the asymptotic states are Fock space states and the fields are completely free in the asymptotic regime of flat spacetime, that is in the region $r \to \infty$ with $u,v=t \pm r$ fixed. Relaxing the assumption of asymptotic decoupling is a hard thing to do. It is certainly unreasonable to consider the full fledged Hamiltonian asymptotically. We can however consider a Hamiltonian that is a good approximation to the original one when the field profiles are well separated. Any interaction involving short wavelengths can thus be disregarded. However, long wavelengths or soft frequencies correspond to long range interactions that should be taken into account. Indeed, it is these frequencies that participate in the resolution of the IR divergences in terms of Faddeev-Kulish (F-K) states \cite{Kulish:1970ut}. The resulting asymptotic Hamiltonian leads to the construction of asymptotic states consisting on Fock space states with clouds of soft photons. The $\mathcal{S}$-matrix between these dressed states are IR finite \cite{Chung:1965zza, Kibble:1968oug, Kibble:1968npb, Kibble:1968lka,Ware:2013zja}. 

In this paper, our objective is to understand the IR finite $\mathcal{S}$-matrix in the flat limit of AdS/CFT. The $\mathcal{S}$-matrix is the only precise observable in flat spacetime. In the presence of gravity, diffeomorphism invariance makes it difficult to define local correlation functions. In AdS, due to the AdS/CFT correspondence, gauge invariant correlation functions can be defined at the AdS boundary as conformal correlators. In flat limit of the AdS/CFT correspondence, the CFT correlation functions are related to the flat space $\mathcal{S}$-matrix, see e.g.\cite{Polchinski:1999ry, Susskind:1998vk, Giddings:1999qu, Giddings:1999jq, Gary:2009mi, Penedones:2010ue,Fitzpatrick:2011ia, Fitzpatrick:2011jn, Fitzpatrick:2010zm, Gary:2009ae,Komatsu:2020sag, vanRees:2022itk, Raju:2012zr, Li:2021snj}. Using the flat limit, the $2 \to 2$ scattering amplitude of massless string states in type IIB superstring theory on $10d$ flat space is evaluated from the $\mathcal{N}=4$ SYM four point functions\cite{Okuda:2010ym}. The analyticity and unitarity of $\mathcal{S}$-matrix from holographic CFT in the flat limit has been elaborated upon in \cite{Fitzpatrick:2011hu, Fitzpatrick:2011dm} and found applications in the $\mathcal{S}$-matrix bootstrap program\cite{Paulos:2016fap, Kruczenski:2022lot}. The flat limit of position space CFT correlation function introduced using HKLL bulk reconstruction prescription introduced in \cite{Hijano:2019qmi, Hijano:2020szl} is another way to formulate holography in flat spacetime. The flat limit relates the bulk scattering states in flat spacetime to boundary CFT operators. Using the map between the scattering states and CFT operators in \cite{Hijano:2020szl} the Weinberg soft theorems are derived from CFT Ward identities. The scattering states for massive scalar particle form a Fock space and in \cite{Hijano:2020szl} the construction of these states are described in AdS/CFT.

Faddeev-Kulish dressings address the IR divergences of the $\mathcal{S}$-matrix in flat spacetime. The dressing can be understood as a Wilson line involving the soft modes of the gauge field \cite{Mandelstam:1962mi, Jakob:1990zi, Choi:2018oel,Choi:2019fuq}. In this paper, we construct the dressing in AdS/CFT in the language of Wilson lines. We then take the flat limit as a large radius of AdS limit. It is common lore that the radius of AdS acts as an IR regulator. As the large limit is taken, IR divergences are expected to reappear unless the asymptotic dynamics of fields is studied appropriately. By constructing dressed fields in AdS, the large radius of AdS limit can be taken while avoiding IR divergences in the flat space observables.

The dressing in a local bulk massive scalar field operator in AdS can be understood as a Wilson line following a path from a bulk point to the boundary. In the soft limit, the scalar field dynamics are coupled to the soft modes of the gauge field. However, the dressed scalar field solves the free Klein-Gordon equation. Therefore, we can reconstruct the dressed scalar field using free HKLL reconstruction method. The interaction of the scalar field with the soft modes of the photon provides the main simplification in our analysis. Now, we can extract the creation mode for the massive scalar field dressed by the Wilson line in terms of dressed CFT operator. The dual CFT operator for the dressed field can be related to the dual CFT operator of the undressed field where the Wilson line end-points are at the boundary. The Wilson line itself can be constructed in terms of current operators in the CFT. The main result of this draft is an expression for the Faddeev-Kulish dressed state in terms of the CFT operators at the boundary. Such expression can potentially be used to compute IR finite S-matrices from flat limits of CFT correlation functions.
\paragraph{\bf Outline of the paper:}~The note is organized as follows. In section \ref{sec:setup} we review very briefly the equivalence between Wilson line dressing and Faddeev-Kulish dressing in flat space and why they are equivalent. In section \ref{a_scalar_gauge} we note down the expressions
for annihilation and creation operators in flat limit of AdS for scalar field and photons, which will be crucial in the latter sections. In section \ref{ads_dressing} we define the soft dressing in AdS. Then in section \ref{dressed_scattering_state} we present the derivation of dressed scattering state in terms of boundary CFT operator and creation and annihilation operator of photon. Finally in section \ref{conclusion} we conclude.

\section{Review: Faddeev-Kulish states as Wilson line dressings}\label{sec:setup}
In the gauge invariant formulation of QED conventional matter fields are dressed by a path dependent Wilson line extending all the way to infinity.\footnote{In this section, we will denote charge by $e$ to avoid notational inconsistency. But in later sections we will again switch to $q$ to denote charge.}
\begin{equation}
\begin{split}
\hat{\Phi}\left(x,\Gamma\right)=\mathcal{P}~\exp\left(ie\int_{\Gamma}^xd\xi^{\mu}~A_{\mu}(\xi)\right)\Phi(x)~.
\end{split}
\end{equation}
In flat space this path dependent dressing describes the Faddeev-Kulish dressing for a particular time-like geodesic path for the asymptotic field \citep{Choi:2018oel}. In this note our goal is to understand how to derive this as a flat limit of AdS/CFT.

In flat space in Lorenz gauge, the Maxwell field equations can be written as
\begin{equation}
\square A_{\mu}=J_{\mu}\, ,
\end{equation}
where, $J_{\mu}$ is the source term. Asymptotically, the solution to this equation consists of a homogeneous solution and a source term
\begin{equation}
A_{\mu} (x) = A^h_{\mu}(x) + \int d^4 x' G_{\text{ret}}(x,x') J_{\mu}(x')\, .
\end{equation}
where $G_{\text{ret}}(x,x^{\prime})$ solves $\Box G_{\text{ret}}(x,x^{\prime})=\delta^{(4)}(x,x^{\prime})$.\\
While source term will contribute by a phase factor, the homogeneous solution will have the following mode expansion
\begin{equation}
\begin{split}
A^{h}_{\mu}(x)=\int\frac{d^3q}{(2\pi)^32\omega_{\vec{q}}}\sum_{\lambda=\pm}\left(\varepsilon^{(\lambda)}_{\mu}\mathbf a^{(\lambda)}_{\vec{q}}e^{iq\cdot x}+{\varepsilon^{(\lambda)}_{\mu}}^{*}{\mathbf a^{(\lambda)}_{\vec{q}}}^{\dagger}e^{-iq\cdot x}\right)
\end{split}
\end{equation}
where, $\omega_{\vec{q}}=|\vec{q}|$, and $\varepsilon^{(\lambda)}_{\mu}$ are the polarization tensor. Choosing the Wilson line path to be the time-like geodesic followed by a semiclassical particle, $\xi^{\mu}=x^{\mu}+\frac{p^{\mu}}{m}\tau$, we have
\begin{equation}\label{W_integral}
\begin{split}
\exp\left(ie\int_{\Gamma}^xd\xi^{\mu}~A_{\mu}(\xi)\right)&=\exp\left\{ie\frac{p^{\mu}}{m}\int_{-\infty}^0d\tau~A_{\mu}^h\left(x^{\nu}+\frac{p^{\nu}}{m}\tau \right)\right\}\\
&=\exp\left\{-e\int\frac{d^3q}{(2\pi)^32\omega_{\vec{q}}}\frac{p^{\mu}}{p\cdot q}\sum_{\lambda=\pm}\left({\varepsilon^{(\lambda)}_{\mu}}^{*}{\mathbf a^{(\lambda)}_{\vec{q}}}^{\dagger}e^{-iq\cdot x}-\varepsilon^{(\lambda)}_{\mu}\mathbf a^{(\lambda)}_{\vec{q}}e^{iq\cdot x}\right)\right\}
\end{split}
\end{equation}
which is the Wilson line dressing of an asymptotic particle with momentum $p^{\mu}$. As we are interested in asymptotic field we will take $x^0\to -\infty$ limit for an asymptotic incoming particle. In that limit the integral in \eqref{W_integral} will have contribution in $q\to 0$ limit and we can replace $e^{\pm iq\cdot x}$ with a function $\phi(p,q)$ which has support in a small neighbourhood of $q\to 0$ with $\phi\to 1$ as $q\to 0$. The resulting expression is the well known formula for the F-K dressing.
\begin{equation}
\begin{split}
\mathcal{W}(p)=\exp\left\{-e\int\frac{d^3q}{(2\pi)^32\omega_{\vec{q}}}\frac{p^{\mu}}{p\cdot q}\phi(p,q)\sum_{\lambda=\pm}\left({\varepsilon^{(\lambda)}_{\mu}}^{*}{\mathbf a^{(\lambda)}_{\vec{q}}}^{\dagger}-\varepsilon^{(\lambda)}_{\mu}\mathbf a^{(\lambda)}_{\vec{q}}\right)\right\}~.
\end{split}
\end{equation}
The expectation is that dressing fields with Wilson lines in the context of AdS/CFT will yield a CFT operator formula for FK dressings upon a flat limit.

\subsection{Why dressing by Wilson line is equivalent to the F-K dressing?}\label{cause_w_dressing}
One may ask behind the motivation for considering a field dressed by a Wilson line. To understand that let us study the equation of motion for the field $\hat{\Phi}$. The scalar electrodynamics action is
\begin{equation}
\begin{split}
S=\int d^{4}x~\sqrt{-g}\left(-D_M\Phi^{\dagger}D^M\phi-m^2\Phi^{\dagger}\Phi-\frac{1}{4}F_{MN}F^{MN}\right) \, ,
\end{split}
\end{equation}
 where $D_M=\partial_M-ieA_{M}$. If we neglect the backreaction of the scalar field $\Phi$ on the field strength $F_{AB}$ we have the equation of motion for the dressed field $\hat{\Phi}$ as \citep{Monica:2017}
\begin{equation}
\label{eqmhatphi1}
\begin{split}
\left(\square-m^2\right)\hat{\Phi}=-ie~\hat{\Phi}~\nabla^M\int_{\Gamma}F_{MA}~dy^A-2ie~\nabla^M\hat{\Phi}~\int_{\Gamma}F_{MA}~dy^A+e^2~\hat{\Phi}~g^{MN}\int_{\Gamma}F_{MA}~dy^A\int_{\Gamma}F_{NB}~dy^B
\end{split}
\end{equation}
Now $F_{AB}~\approx \mathcal{O}(q)$ and if we consider the soft limit (i.e $q\to 0$ limit) then we can neglect the right hand side of the equation \eqref{eqmhatphi1} and eventually we have
\begin{equation}
\begin{split}
\left(\square-m^2\right)\hat{\Phi}=\mathcal{O}(q)
\end{split}
\end{equation}
Hence, in the soft limit the dressed field actually behaves like a free field. IR divergences are a consequence of the non-trivial interactions among fields. In particular interactions involving long range forces. We can expect states corresponding to free fields will not have the problem of bringing IR divergences.

\section{Free scalar and gauge field in AdS and it's flat space limit}\label{a_scalar_gauge}
In this note our main goal is to construct these dressed state as a flat limit from AdS. For that let us first recap HKLL bulk field reconstruction in global AdS.\\

We start with the $\text{AdS}_4$ global coordinate where the line element reads
\begin{equation}
\begin{split}
ds^2=\frac{L^2}{\cos^2\rho}\left(-d\tau^2+d\rho^2+\sin^2\rho~d\Omega^2_2\right)
\end{split}
\end{equation}
Where $\rho$ is the radial coordinate and the boundary of the space is located at $\rho=\frac{\pi}{2}$. The coordinates parametrizes the CFT are $x\equiv \left(\tau,\Omega_2\right)$ and $L$ is  the AdS radius.\par 

Now let us try to understand how to perform a flat limit in this space. We need to take AdS radius $L$ to be large with $\tau=t/L$ and $\rho=r/L$ such that
\begin{equation}
\begin{split}
ds^2~\xrightarrow[L\to\infty]~-dt^2+dr^2+r^2~d\Omega^2_2
\end{split}
\end{equation}
In this global coordinate we can think of the the space-time a cylinder and and taking a flat limit implies zooming into the centre of the cylinder where geometry is locally flat.
\subsection{Free scalar field in AdS}\label{free_field_O}
In this note we will use HKLL bulk operator reconstruction in global AdS and discuss explicitly the flat limit to obtain the flat space creation and annihilation operators in terms of CFT operator smeared along time-like direction.\\

We consider the scattering region at $\tau=0$, at the centre of the AdS. Now if we consider the assumption of asymptotic decoupling then before and after the scattering region the scalar field $\Phi$ must obey free Klein-Gordon equation on AdS
\begin{equation}
\begin{split}
\left(\square-m^2\right)\Phi=0,~~\text{with}~~m^2L^2=\Delta\left(\Delta-3\right)
\end{split}
\end{equation}
with the following fall-off behaviour
\begin{equation}
\begin{split}
\Phi(\rho,x)~\xrightarrow[\rho\to\frac{\pi}{2}]~\left(\cos\rho\right)^{\Delta}\mathcal{O}(x)
\end{split}
\end{equation}
After reconstructing these operators in HKLL bulk reconstruction method we need to take a flat limit to extract the creation and annihilation operators. For massive field this reads \citep{Hijano:2020szl}
\begin{equation}
\begin{split}
\sqrt{2\omega_{\vec{p}}}~{a}^{\dagger}_{\omega_{\vec{p}}}&={C}\left(L,m,|\vec{p}|\right)\int_0^{\pi} d\tau e^{-i\omega_{\vec{p}} L\left[\tau-\frac{\pi}{2}-\frac{i}{2}\log\left(\frac{\omega_{\vec{p}}+m}{\omega_{\vec{p}}-m}\right)\right]}\mathcal{O}\left(\tau,\hat{p}\right)
\end{split}
\end{equation}
where, $\hat{p}$ is the angle of the momentum $\vec{p}$ and $
{C}=\frac{1}{2\pi} \Bigg(\frac{mL}{\pi^3}\Bigg)^{\frac{1}{4}} \Bigg(\frac{2m}{i |\vec{p}|}\Bigg)^{mL+\frac{1}{2}}~ L \,.
$ This integral will be dominated by a window of size of $\mathcal{O}\left(\frac{1}{L}\right)$ at $\text{Re}(\tau)=\pm\frac{\pi}{2}$ and $\text{Im}(\tau)=\frac{1}{2}\log\left(\frac{\omega_{\vec{p}}+m}{\omega_{\vec{p}}-m}\right)$. This has been denoted by the caps in figure \ref{fig:HKLL}. 
\subsection{Free $U(1)$ gauge field in AdS}
Following \citep{Hijano:2020szl} in this section we review the construction of photon scattering states in flat limit of AdS. Again, by virtue of asymptotic decoupling, we will assume that photons obey the free Maxwell equation at early and late times. There are two possibilities to quantize the bulk gauge field, which are specified by either electric or magnetic boundary conditions. In this note we will consider the magnetic boundary condition, which fixes the magnetic field at the boundary to zero. In this case the gauge is dual to a current operator in the CFT. We thus look for a field operator with boundary condition
\begin{equation}
\begin{split}
{\mathcal{A}_{\mu}}(\rho,x) \xrightarrow[\rho\to\frac{\pi}{2}] ~ \cos\rho \, j_{\mu}(x)\, .
\end{split}
\end{equation}
A free $U(1)$ gauge field operator in Minkowaski space can be written in the following form.
\begin{equation}\label{aaA}
\begin{split}
{\cal A}_{\mu}(x)=\int \frac{d^3\vec{q}}{(2\pi)^3}\frac{1}{\sqrt{2\omega_{\vec{q}}}}\sum_{\lambda=\pm}\left(\varepsilon^{(\lambda)}_{\mu}\mathbf a^{(\lambda)}_{\vec{q}}e^{iq\cdot x}+{\varepsilon^{(\lambda)}_{\mu}}^{*}{\mathbf a^{(\lambda)}_{\vec{q}}}^{\dagger}e^{-iq\cdot x}\right),~~x\in \text{Mink}_{3+1}\, ,
\end{split}
\end{equation}
where, $\varepsilon^{(\lambda)}_{\mu}$ are the null polarization vector with $\lambda=\pm$ denote the two different polarization and it is normalized such that $\varepsilon^{(+)}_{\mu}\varepsilon^{(-),\mu}=1$. We will also work in Lorenz gauge such that $\varepsilon_{\mu}q^{\mu}=0$.\par
\begin{figure}[]
\centering
\begin{tikzpicture}[scale=0.7]
\shade[top color=white, bottom color=black!45!white,opacity=0.75]  (5,2) arc (0:180:3 and 1)  --  (5,2) arc (0:-180:3 and 3)   -- cycle;
\shade[top color=white, bottom color=gray,opacity=0.75]  (-1,2) arc (180:360:3 and 1)   --  (5,2) arc (0:-180:3 and 3)   -- cycle;
\draw[black,very thick] (5,2) arc (0:-180:3 and 3) node[pos=0.5,below]{$\partial{\cal M}_-$};

\draw[gray, thick,dotted] (2,2) -- (2,7) node [pos=1,above=2]{{$ $}} node [pos=0,below=2]{{$ $}};
\draw[blue,very thick,dashed,opacity=0,name path=TOPBD2](5,1.7) arc (0:180:3 and 1);
\draw[blue,very thick,dashed,name path=LOWBD](5,2) arc (0:180:3 and 1);
\draw[blue,very thick,dashed,opacity=0,name path=TOPBD](5,2.3) arc (0:180:3 and 1);
\tikzfillbetween[of=TOPBD and TOPBD2]{blue, opacity=0.5};

\draw[blue,very thick,opacity=0, name path=TOPB2](-1,1.7) arc (180:360:3 and 1);
\draw[blue,very thick, name path=LOWB](-1,2) arc (180:360:3 and 1);
\draw[blue,very thick,opacity=0, name path=TOPB](-1,2.3) arc (180:360:3 and 1);
\tikzfillbetween[of=TOPB and TOPB2]{blue, opacity=0.5};

\draw[black,very thick] (-1,2) -- (-1,7) node [pos=0,left]{\color{blue} $\tilde{\cal I}^-$} node [pos=1,left]{\color{red} $\tilde{\cal I}^+$};
\draw[black,very thick] (5,2) -- (5,7) node [pos=1,right]{$\tau={{\pi}\over 2}$} node [pos=0,right]{$\tau=-{{\pi}\over 2}$};

\draw[red,very thick,dashed,opacity=0,name path=TOPRD2](5,7.3) arc (0:180:3 and 1);
\draw[red,very thick,dashed,name path=TOPRD](5,7) arc (0:180:3 and 1);
\draw[red,very thick,dashed,opacity=0,name path=LOWRD](5,6.7) arc (0:180:3 and 1);
\tikzfillbetween[of=TOPRD2 and LOWRD]{red, opacity=0.5};

\draw[black,very thick] (5,7) arc (0:180:3 and 3) node[pos=0.5,above]{$\partial{\cal M}_+$};
\shade[top color=gray,bottom color=white,opacity=0.75]  (-1,7) arc (180:360:3 and 1)  --  (5,7) arc (0:180:3 and 3)  -- cycle;

\draw[red,very thick,opacity=0,name path=TOPR2](-1,7.3) arc (180:360:3 and 1);
\draw[red,very thick,name path=TOPR](-1,7) arc (180:360:3 and 1);
\draw[red,very thick,opacity=0,name path=LOWR](-1,6.7) arc (180:360:3 and 1);
\tikzfillbetween[of=TOPR2 and LOWR]{red, opacity=0.5};

\draw[cyan, very thick] (3.5,5.82) -- (3.5,5.82+0.6);
\draw[cyan, very thick, snake it,->] (2,4) -- (3,5.5);

\end{tikzpicture}
\caption{Pictorial representation of the map between  the asymptotic regions of flat space and the boundary of AdS. The euclidean domes $\partial \mathcal{M}_{\pm}$ are analytic continuation of the boundary CFT in the imaginary global time direction playing the role of future/past infinity $i^{\pm}$. $\widetilde{\mathcal{I}}^{\pm}$ are future/past null infinity. This figure is taken from \citep{Hijano:2020szl}.}\label{fig:HKLL}
\end{figure}
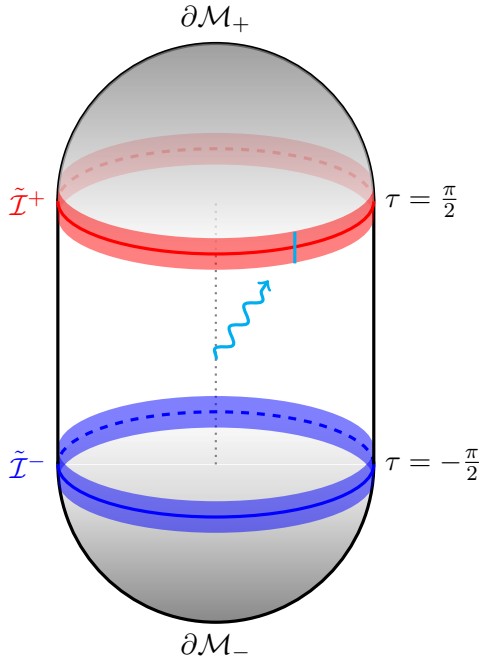

 
In Lorenz gauge we can reconstruct the gauge field explicitly using HKLL and then taking a flat limit and using \eqref{aaA} we can derive the creation/annihilation modes explicitly. This has been done in \citep{Hijano:2020szl}. We have
\begin{equation}\label{aaj}
\begin{split}
&\sqrt{2\omega_{\vec{q}}}~\mathbf a^{\dagger (-)}_{\vec{q}}=\frac{-1}{4\omega_{\vec{q}}}\frac{1+z_q\bar{z}_q}{\sqrt{2}}\int d\tau^{\prime}~e^{i\omega_{\vec{q}}L\left(\frac{\pi}{2}-\tau^{\prime}\right)}\int d^2z^{\prime}\frac{1}{\left(z_q-z^{\prime}\right)^2}j^+_{\bar{z}^{\prime}}(\tau^{\prime},z^{\prime},\bar{z}^{\prime})\\
&\sqrt{2\omega_{\vec{q}}}~ \mathbf a^{\dagger (+)}_{\vec{q}}=\frac{-1}{4\omega_{\vec{q}}}\frac{1+z_q\bar{z}_q}{\sqrt{2}}\int d\tau^{\prime}~e^{i\omega_{\vec{q}}L\left(\frac{\pi}{2}-\tau^{\prime}\right)}\int d^2z^{\prime}\frac{1}{\left(\bar{z}_q-\bar{z}^{\prime}\right)^2}j^+_{{z}^{\prime}}(\tau^{\prime},z^{\prime},\bar{z}^{\prime})\\
&\sqrt{2 \omega_{\vec{q}}}~ \mathbf a_{\vec{q}}^{(-)}
=-\frac{1}{4\omega_{\vec{q}}}\frac{1+z_{q}\bar{z}_{q}}{\sqrt{2}}\int d\tau^{\prime} e^{-i \omega_{\vec{q}}L(\frac{\pi}{2}-\tau^{\prime})} \int d^{2}z^{\prime} \frac{1}{(z_q-z^{\prime})^2}  j^{-}_{\bar{z}^{\prime}}(\tau^{\prime},z^{\prime},\bar{z}^{\prime})~~~\\
&\sqrt{2 \omega_{\vec{q}}}~ \mathbf a_{\vec{q}}^{(+)}
=-\frac{1}{4\omega_{\vec{q}}}\frac{1+z_{q}\bar{z}_{q}}{\sqrt{2}}\int d\tau^{\prime} e^{-i \omega_{\vec{q}}L(\frac{\pi}{2}-\tau^{\prime})} \int d^{2}z^{\prime} \frac{1}{(\bar{z}_q-\bar{z}^{\prime})^2}  j^{-}_{z^{\prime}}(\tau^{\prime},z^{\prime},\bar{z}^{\prime})
\end{split}
\end{equation}
where $(z,\bar z)$ are the complex coordinate on the sphere and $\mathbf a^{\dagger(\pm)}_{\vec{q}}, \mathbf a^{(\pm)}_{\vec{q}}$ are the photon creation and annihilation operators respectively. These expression tells us that the integral will be dominated by a window of size of $\mathcal{O}\left(\frac{1}{L}\right)$ around the region $\tau=+ \frac{\pi}{2}$ for outgoing particle and $\tau=-\frac{\pi}{2}$ for incoming particle. A pictorial representation of these formulas can be found in figure \ref{fig:HKLL}.

\section{Soft dressing in AdS}\label{ads_dressing}
In this section we will try to define the dressing in AdS that could correspond to the F-K state after taking a flat limit.\\

In \citep{Hijano:2020szl} the authors have studied fields that are completely free in the asymptotic regime of Minkowaski space. That is the fields are free in the region $r\to\infty$ with $u=t-r$ fixed. Now relaxing the assumption of asymptotic decoupling involves including interaction in the definition of scattering states. Then instead of studying free fields we could study fields that interact with the long wavelength modes of the electro-magnetic field. We could then think of reconstructing such interacting fields in AdS/CFT and taking a flat limit described in previous section. Then one can expect that the resulting scattering state correspond to F-K state that are usual Fock space states dressed by clouds of photons.
\subsection{Defining the soft dressing in AdS}
Now the question is how to define the dressing in AdS. Taking the motivation from flat space described in \ref{cause_w_dressing} we will define the dressing in the bulk of AdS and then consider reconstruction of those fields in the bulk of AdS by HKLL bulk reconstruction method.\\

In flat space, the soft limit is attained by the frequency of the photon being zero. Soft frequencies are the ones participating in long range forces, and so they are the only ones that should be considered when constructing asymptotic fields in the flat limit. In flat space we can take minimum momentum of photon to be zero, $q\to 0$. But in AdS this is not possible because minimum frequency would be inversely proportional to the size of the space. Hence we can think that in the soft limit the field strength $F_{AB}$ would be $\mathcal{O}\left(\frac{1}{L}\right)$. Taking this as the input we can define the dressing in AdS in the following way.\\

Let us consider a free scalar field $\Phi$ dual to an operator $\mathcal{O}$ and a gauge field $\mathcal{A}$ dual to an operator $j$. Now we will define the dressing by the following non-local but gauge invariant bulk operator
\begin{equation}
\Tilde\Phi(y)=U(x,y)\Phi(y)
\end{equation}
along with $F_{AB}=\mathcal{O}\left(\frac{1}{L}\right)$.\\

The point $y$ is a point in the bulk of AdS and the point $x$ is a point in the boundary of AdS. $U(x,y)$ is a Wilson line constructed from the gauge field. Explicitly we have
\begin{equation}
U(x,y)=\mathcal{P}\Bigg\{e^{iq\int_{x}^{y}\mathcal{A}_M dx^M}\Bigg\}
\end{equation}
Now the bulk field $\Tilde\Phi$ is gauge invariant by definition but as a trade-off we have introduced a nonlocality because the field now depend on a boundary point $x$.\par 

If we neglect the backreaction of the scalar field $\Phi$ on the field strength $F_{AB}$ we have the equation of motion for the dressed field $\Tilde{\Phi}$ as \citep{Monica:2017}
\begin{equation}
\label{eqmhatphi}
\begin{split}
\left(\square-m^2\right)\Tilde{\Phi}=-iq~\Tilde{\Phi}~\nabla^M\int_{\Gamma}F_{MA}~dy^A-2iq~\nabla^M\Tilde{\Phi}~\int_{\Gamma}F_{MA}~dy^A+q^2~\Tilde{\Phi}~g^{MN}\int_{\Gamma}F_{MA}~dy^A\int_{\Gamma}F_{NB}~dy^B
\end{split}
\end{equation}
In the soft limit as we have $F_{AB}=\mathcal{O}\left(\frac{1}{L}\right)$, then the equation of motion for the field $\Tilde\Phi$, becomes
\begin{equation}
\begin{split}
\left(\square-m^2\right)\Tilde{\Phi}=\mathcal{O}\left(\frac{1}{L}\right)
\end{split}
\end{equation}
Hence, in the kinematic regime we are interested the field $\Tilde{\Phi}$ obeys \textbf{free} Klein-Gordon equation. This means we can actually reconstruct $\Tilde{\Phi}$ by using free HKLL, without worrying about interactions. Of course, the operator at the boundary is now different. Instead of having usual operator $\mathcal{O}$ dual to free bulk field $\Phi$, now we have a non-local operator $\tilde{O}$.
\subsection{The dual operator of the AdS dressing in CFT}
\begin{figure}[H]	
\centering
\includegraphics[width=0.25\linewidth]{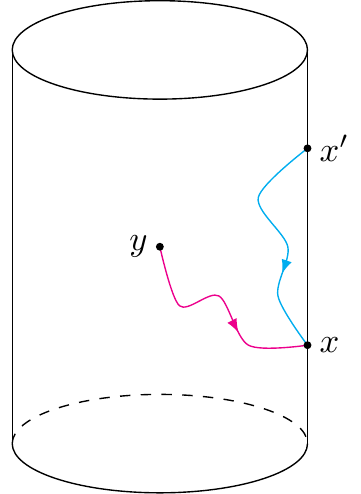}
\caption{Wilson line: the magenta line is the Wilson line connecting bulk point $y$ to boundary point $x$. The cyan line is the Wilson line connecting two boundary points $x^{\prime}$ and $x$.}
\label{fig:wilsonline}
\end{figure}

Let us now determine the dual operator to the dressed field $\Tilde\Phi$
\begin{equation}
\Tilde\Phi(y)=\mathcal{P}\Bigg\{e^{iq\int_{x}^{y}\mathcal{A}_M dx^M}\Bigg\}\Phi(y)~~~.
\end{equation}
Since,\footnote{where $a$ are coordinates at the boundary.}
\begin{equation}
\begin{split}
\Phi(\rho,x)~&\xrightarrow[\rho\to\frac{\pi}{2}]~\left(\cos\rho\right)^{\Delta}\mathcal{O}(x)\\
\mathcal{A}_a\left(\rho,x\right)~&\xrightarrow[\rho\to\frac{\pi}{2}]~j_a~\cos\rho
\end{split}
\end{equation}
and the line element in the bulk varies with the boundary line element with a factor of $\frac{1}{\cos\rho}$, the dual operator should be 
\begin{equation}
\tilde{O}(x^{\prime})=\mathcal{P}\Bigg\{e^{i q \int_{x}^{x^{\prime}} dx^a j_a}\Bigg\} \mathcal{O}(x^{\prime})~.
\end{equation}
As the end-points of this Wilson line are at the boundary, we can thought this Wilson line as an expression in the boundary. 
In fig.\ref{fig:wilsonline} we draw two Wilson lines, one representing the Wilson line in AdS joining bulk point to boundary point and the other is the Wilson line joining two boundary points. 
\section{The flat space scattering state for the dressed operator}\label{dressed_scattering_state}
\subsection{The dressed scattering state in terms of CFT operators}
In flat space the decomposition of a scalar field $\Phi$ can be done at early and late times. Hence, there exists two sets of creation and annihilation operators -one for early and the other for the late time. This means the ``in'' and ``out'' scattering state can be built using the formulas
\begin{equation}
\begin{split}
&\left |\vec{k},in\right>=\sqrt{2\omega_{\vec{k}}}~a^{\dagger}_{in,\vec{k}}\left |0\right>\\
&\left<out,\vec{p}\right|=\sqrt{2\omega_{\vec{p}}}~\left<0\right|a_{out,\vec{p}}
\end{split}
\end{equation}
The $\mathcal{S}$ matrix is computed as an overlap between ``in'' and ``out'' states
\begin{equation}
\mathcal{S}_{fi}=\left<out|in\right>
\end{equation}
In this way the $1\rightarrow 1$ $\mathcal{S}$-matrix element is normalized as
\begin{equation}
\left<k_1|\mathcal{S}|k_2\right>=2\omega_{\vec{k}}~\delta^{(3)}\left(\vec{k_1}-\vec{k_2}\right)
\end{equation}
Now let us discuss how to construct the dressed ``in'' and ``out'' state in flat space. As we have reviewed in \ref{free_field_O} for a free scalar field $\Phi$ in the bulk of AdS with the dual operator $\mathcal{O}$, the creation operator in flat space reads
\begin{equation}
\begin{split}
\sqrt{2\omega_{\vec{p}}}~{a}^{\dagger}_{\omega_{\vec{p}}}&=\tilde{C}\left(L,m,|\vec{p}|\right)\int_0^{\pi} d\tau e^{-i\omega_{\vec{p}} L\left[\tau-\frac{\pi}{2}-\frac{i}{2}\log\left(\frac{\omega_{\vec{p}}+m}{\omega_{\vec{p}}-m}\right)\right]}\mathcal{O}\left(\tau,\hat{p}\right)
\end{split}
\end{equation}
Now instead of a free scalar field $\Phi$ we have a free field $\Tilde\Phi$ with dual operator $\Tilde{\mathcal{O}}$ . Hence, the creation modes extracted from the position space scalar field operator as
\begin{equation}
\begin{split}
\sqrt{2\omega_{\vec{p}}}~\Tilde{a}^{\dagger}_{\omega_{\vec{p}}}&=\tilde{C}\left(L,m,|\vec{p}|\right)\int_0^{\pi} d\tau e^{-i\omega_{\vec{p}} L\left[\tau-\frac{\pi}{2}-\frac{i}{2}\log\left(\frac{\omega_{\vec{p}}+m}{\omega_{\vec{p}}-m}\right)\right]}\Tilde{\mathcal{O}}\left(\tau,\hat{p}\right)
\end{split}
\end{equation}
This operator when act on the CFT vacuum will give the IR finite $\mathcal{S}$ matrix in CFT.\par 

Now using the expression for $\Tilde{\mathcal{O}}$ we can write the dressed creation operator as
\begin{equation}
\begin{split}
\sqrt{2\omega_{\vec{p}}}~\Tilde{a}^{\dagger}_{\omega_{\vec{p}}}&=\tilde{C}\left(L,m,|\vec{p}|\right)\int_0^{\pi} d\tau e^{-i\omega_{\vec{p}} L\left[\tau-\frac{\pi}{2}-\frac{i}{2}\log\left(\frac{\omega_{\vec{p}}+m}{\omega_{\vec{p}}-m}\right)\right]}\exp\left(iq\int_{\Gamma(\tau,\hat{p})}j_a~dx^a\right){\mathcal{O}}\left(\tau,\hat{p}\right)
\end{split}
\end{equation}
As in the kinematic regime we are interested we can neglect $F_{AB}$, which implies that the Wilson line is path independent and we can choose any path. We will choose
\begin{equation}\label{the_path}
\begin{split}
&\Gamma\left(\tau,\hat{p}\right)~:~x\rightarrow x_0\\
&x^a_0~:~x^{\tau}_0=0,\hat{\Omega}_0=\hat{p}\\
&x^a~:~x^{\tau}=\tau,\hat{\Omega}_0=\hat{p}\\
\end{split}
\end{equation}
Then,
\begin{equation}
\begin{split}
\exp\left(iq\int_{\Gamma(\tau,\hat{p})}j_a~dx^a\right)&=\exp\left(iq\int_0^{\tau}j_{\tau^{\prime}}d\tau^{\prime}\right)\\
&=1+iq\int_0^{\tau}j_{\tau^{\prime}}d\tau^{\prime}+\mathcal{O}(q^2)\\
&=1+q\sum_m \frac{1}{\omega_m}\left[e^{i\omega_m\tau}-1\right]\hat{X}_m+\mathcal{O}(q^2)\\
\end{split}
\end{equation}
Where we have used the form of the CFT operator $j_a$, which is
\begin{equation}
\begin{split}
j_{\tau^{\prime}}\left(\tau^{\prime},\hat{p}\right)=\sum_m e^{i\omega_m\tau^{\prime}}\hat{X}_m(\hat{p})+h\cdot c
\end{split}
\end{equation}
where,
\begin{equation}
\omega_m=1+l+2\kappa~~\text{or}~~2+l+2\kappa \, .
\end{equation}\\
Upon taking the flat limit $L\to \infty$, we need to specify what happens to the integers $\kappa$. A possibility is to interpret any mode with $\kappa \sim {\cal O}(1)$ as a soft mode, as their frequency tends to zero $\omega_m/L \rightarrow 0$. This is however not what we usually consider a soft field. We usually take the limit from finite momentum to zero, which means we first write $\kappa  = k L$, then take the large $L$ limit, and finally we take the $k\rightarrow 0$ limit. This is the limit we will consider here. It is however interesting to wonder what the physical meaning of modes with finite $\kappa$ is. All in all, we will define the soft limit as $\frac{\omega_m}{L}=k$ with $k\rightarrow 0$ after $L\to \infty$.\\
Now as $\omega_m=1+l+2\kappa=kL$, for such large value of $\kappa$ the sum over integers turn into integral over $k$.
\begin{equation}
\sum_{m}\rightarrow \frac{L}{2}\int dk
\end{equation}
So, we have
$$q\sum_m \frac{1}{\omega_m}\left[e^{i\omega_m\tau}-1\right]\hat{X}_m(\hat{p})=\frac{q}{2}\int \frac{dk}{k}\left[e^{iLk\tau}-1\right]\hat{X}_k(\hat{p})$$
So, up to $\mathcal{O}(q)$ the dressed creation operator can be written as
\begin{equation}
\begin{split}
\sqrt{2\omega_{\vec{p}}}~\Tilde{a}^{\dagger}_{\omega_{\vec{p}}}=&\tilde{C}\left(L,m,|\vec{p}|\right)\int_0^{\pi} d\tau e^{-i\omega_{\vec{p}} L\left[\tau-\frac{\pi}{2}-\frac{i}{2}\log\left(\frac{\omega_{\vec{p}}+m}{\omega_{\vec{p}}-m}\right)\right]}{\mathcal{O}}\left(\tau,\hat{p}\right)\\
&+\frac{q}{2}\tilde{C}\left(L,m,|\vec{p}|\right)\int_0^{\pi}d\tau\int \frac{dk}{k}e^{-i\omega_{\vec{p}} L\left[\tau-\frac{\pi}{2}-\frac{i}{2}\log\left(\frac{\omega_{\vec{p}}+m}{\omega_{\vec{p}}-m}\right)\right]}\left[e^{iLk\tau}-1\right]\hat{X}_k(\hat{p})~{\mathcal{O}}\left(\tau,\hat{p}\right)+\mathcal{O}(q^2)
\end{split}
\end{equation}
This formula can potentially be used to compute IR finite S-matrices. Firstly, one would have to evaluate the CFT correlator in an integral representation like a Mellin amplitude, and then perform the smearing integrals in a saddle point approximation involving the large $L$ limit, and finally, the soft $k$ limit.

\subsection{The dressed state in terms of creation and annihilation operator of scalar field and photon}\label{dressedaa}
From the previous section we have the dressed creation operator as
\begin{equation}
\begin{split}
\sqrt{2\omega_{\vec{p}}}\tilde{a}^{\dagger}_{\omega_{\vec{p}}}&=\tilde{C}\int d\tau e^{-i\omega_{\vec{p}} L\left[\tau-\frac{\pi}{2}-\frac{i}{2}\log\left(\frac{\omega_{\vec{p}}+m}{\omega_{\vec{p}}-m}\right)\right]}e^{iq\int_{\Gamma(\tau,\hat{p})}j_adx^a}\mathcal{O}\left(\tau,\hat{p}\right)\\
\end{split}
\end{equation}
Now we will analyze the expression in the following way. At first we will introduce a delta function inside the integral and then use the following expression of the Dirac delta function
\begin{equation}
\begin{split}
\delta(\tau-\tau^{\prime})=\frac{L}{2\pi}\int d\Delta_{\vec{p}}~e^{i\Delta_{\vec{p}}L(\tau-\tau^{\prime})}
\end{split}
\end{equation}
Then we can write the dressed operator as
\begin{equation}
\begin{split}
\sqrt{2\omega_{\vec{p}}}\tilde{a}^{\dagger}_{\omega_{\vec{p}}}&=\tilde{C}\int d\tau \left[\int d\tau^{\prime}\delta\left(\tau-\tau^{\prime}\right)\right]e^{-i\omega_{\vec{p}} L\left[\tau-\frac{\pi}{2}-\frac{i}{2}\log\left(\frac{\omega_{\vec{p}}+m}{\omega_{\vec{p}}-m}\right)\right]}e^{iq\int_{\Gamma(\tau,\hat{p})}j_adx^a}\mathcal{O}\left(\tau,\hat{p}\right)\\
&=\frac{\tilde{C}L}{2\pi}\int d\tau d\tau^{\prime}d\Delta_{\vec{p}}~e^{i\Delta_{\vec{p}} L\left(\tau-\tau^{\prime}\right)}
e^{-i\omega_{\vec{p}} L\left[\tau-\frac{\pi}{2}-\frac{i}{2}\log\left(\frac{\omega_{\vec{p}}+m}{\omega_{\vec{p}}-m}\right)\right]}e^{iq\int_{\Gamma(\tau,\hat{p})}j_adx^a}\mathcal{O}\left(\tau^{\prime},\hat{p}\right)\\
\end{split}
\end{equation}
Now after rearranging the expression a little bit we have the following expression
\begin{equation}
\begin{split}
\sqrt{2\omega_{\vec{p}}}\tilde{a}^{\dagger}_{\omega_{\vec{p}}}=&\frac{\tilde{C}L}{2\pi}\int d\Delta_{\vec{p}} ~e^{-i\Delta_{\vec{p}} L\left[\frac{\pi}{2}+\frac{i}{2}\log\left(\frac{\Delta_{\vec{p}}+m}{\Delta_{\vec{p}}-m}\right)\right]}e^{i\omega_{\vec{p}} L\left[\frac{\pi}{2}+\frac{i}{2}\log\left(\frac{\omega_{\vec{p}}+m}{\omega_{\vec{p}}-m}\right)\right]}\int d\tau e^{iL\tau\left(\Delta_{\vec{p}}-\omega_{\vec{p}}\right)}e^{iq\int_{\Gamma(\tau,\hat{p})}j_adx^a}\\
&\frac{1}{C}~C~\int d\tau^{\prime}e^{-i\Delta_{\vec{p}} L\left[\tau^{\prime}-\frac{\pi}{2}-\frac{i}{2}\log\left(\frac{\Delta_{\vec{p}}+m}{\Delta_{\vec{p}}-m}\right)\right]}\mathcal{O}\left(\tau^{\prime},\hat{p}\right)\\
\end{split}
\end{equation}
Now we use the definition of the creation operator
\begin{equation}
\sqrt{2\Delta_{\vec{p}}}~{a}^{\dagger}_{\Delta_{\vec{p}}}=C\int d\tau^{\prime}e^{-i\Delta_{\vec{p}} L\left[\tau^{\prime}-\frac{\pi}{2}-\frac{i}{2}\log\left(\frac{\Delta_{\vec{p}}+m}{\Delta_{\vec{p}}-m}\right)\right]}\mathcal{O}\left(\tau^{\prime},\hat{p}\right)
\end{equation}
Then the dressed creation operator will have the following form
\begin{equation}\label{a_dagger_f}
\begin{split}
\sqrt{2\omega_{\vec{p}}}\tilde{a}^{\dagger}_{\omega_{\vec{p}}}=\frac{\tilde{C}L}{2\pi}&\int d\Delta_{\vec{p}} ~e^{-i\Delta_{\vec{p}} L\left[\frac{\pi}{2}+\frac{i}{2}\log\left(\frac{\Delta_{\vec{p}}+m}{\Delta_{\vec{p}}-m}\right)\right]}e^{i\omega_{\vec{p}} L\left[\frac{\pi}{2}+\frac{i}{2}\log\left(\frac{\omega_{\vec{p}}+m}{\omega_{\vec{p}}-m}\right)\right]}\\
&
\int d\tau e^{iL\tau\left(\Delta_{\vec{p}}-\omega_{\vec{p}}\right)}e^{iq\int_{\Gamma(\tau,\hat{p})}j_adx^a}\frac{\sqrt{2\Delta_{\vec{p}}}}{C}~{a}^{\dagger}_{\Delta_{\vec{p}}}\\
\end{split}
\end{equation}
Now the above expression has been written in terms of creation/annihilation operator of the scalar field and an operator constructed from the CFT current $j_a$, dual to the gauge field. So to write this dressed creation operator totally in terms of creation/annihilation operator the remaining task is to write the expression $e^{iq\int_{\Gamma(\tau,\hat{p})}j_adx^a}$ in terms of creation/annihilation of photon. To do that first note from the previous section that for a particular path of \eqref{the_path} we have
\begin{equation}\label{exp_I_j}
\begin{split}
\exp\left(iq\int_{\Gamma(\tau,\hat{p})}j_a~dx^a\right)&=\exp\left(iq\int_0^{\tau}j_{\tau^{\prime}}d\tau^{\prime}\right)=\exp\left(iq\int_0^{\tau}\left(j^+_{\tau^{\prime}}+j^-_{\tau^{\prime}}\right)d\tau^{\prime}\right)
\end{split}
\end{equation}
where, $+$ and $-$ symbol are due to the positive and negative frequency modes of the gauge field.\\
Then we use the conservation equation $\partial_{\mu}j^{\mu}=0$ to write $j_{\tau^{\prime}}$ in terms of creation/annihilation operators as follows
\begin{equation}
\begin{split}
&\partial_{\mu}j^{\mu}=0\\
\implies&\partial_{\tilde\tau}j^{\tilde\tau}+\partial_{z}j^{z}+\partial_{\bar{z}}j^{\bar{z}}=0\\
\implies &\partial_{\tilde\tau}j_{\tilde\tau}=\frac{1}{2}(1+z\bar{z})^2(\partial_{z}j_{\bar{z}}+\partial_{\bar{z}}j_{z})+(1+z\bar{z})\left(zj_{z}+\bar{z}j_{\bar{z}}\right)
\end{split}
\end{equation}
,where we have used the fact that the metric on the sphere is $ds^2=\frac{4}{(1+z\bar{z})^2}dz~d\bar{z}$.
Then integrating out the above expression we have the following expression for $j_{\tau^{\prime}}$.
\begin{equation}\label{j_tau_a}
\begin{split}
j_{\tau^{\prime}}^{\pm}=\int_{0}^{\tau^{\prime}}d{\tilde\tau}\Bigg(\frac{1}{2}(1+z\bar{z})^2(\partial_{z}j_{\bar{z}}^{\pm}+\partial_{\bar{z}}j_{z}^{\pm})+(1+z\bar{z})\left(zj_{z}^{\pm}+\bar{z}j_{\bar{z}}^{\pm}\right)\Bigg)
\end{split}
\end{equation}
Now to write the above expression in terms of creation/annihilation operator of photon we need to invert the expression of \eqref{aaj}, which have the following expressions\footnote{The derivation of the expressions has been written in \ref{appendixA}.}
\begin{equation}
\begin{split}
\partial_{z_q}j^{+}_{\bar{z}_q}(\tau,z_q,\bar{z}_q)&=\frac{L}{(2\pi)^2}\int d\omega_{\vec{q}}~e^{i\omega_{\vec{q}}L\left(\tau-\pi/2\right)}\partial_{\bar{z}_q}\left(\frac{8\omega_{\vec{q}}\sqrt{\omega_{\vec{q}}}~\mathbf a^{\dagger (-)}_{\vec{q}}}{1+z_q\bar{z}_q}\right)\\
j^{+}_{\bar{w}}(\tau,w,\bar{w})&=\frac{-L}{(2\pi)^3}\int d\omega_{\vec{q}}~e^{i\omega_{\vec{q}}L\left(\tau-\pi/2\right)}\int d^2z_q\frac{1}{(\bar{z}_q-\bar{w})}\partial_{\bar{z}_q}\left(\frac{8\omega_{\vec{q}}\sqrt{\omega_{\vec{q}}}~\mathbf a^{\dagger (-)}_{\vec{q}}}{1+z_q\bar{z}_q}\right)\\
\partial_{\bar{z}_q}j^{+}_{{z}_q}(\tau,z_q,\bar{z}_q)&=\frac{L}{(2\pi)^2}\int d\omega_{\vec{q}}~e^{i\omega_{\vec{q}}L\left(\tau-\pi/2\right)}\partial_{{z}_q}\left(\frac{8\omega_{\vec{q}}\sqrt{\omega_{\vec{q}}}~\mathbf a^{\dagger (+)}_{\vec{q}}}{1+z_q\bar{z}_q}\right)\\
j^{+}_{{w}}(\tau,w,\bar{w})&=\frac{-L}{(2\pi)^3}\int d\omega_{\vec{q}}~e^{i\omega_{\vec{q}}L\left(\tau-\pi/2\right)}\int d^2z_q\frac{1}{({z}_q-{w})}\partial_{{z}_q}\left(\frac{8\omega_{\vec{q}}\sqrt{\omega_{\vec{q}}}~\mathbf a^{\dagger (+)}_{\vec{q}}}{1+z_q\bar{z}_q}\right)\\
\partial_{{z}_q}j^{-}_{\bar{z}_q}(\tau,z_q,\bar{z}_q)&=\frac{L}{(2\pi)^2}\int d\omega_{\vec{q}}~e^{-i\omega_{\vec{q}}L\left(\tau-\pi/2\right)}\partial_{\bar{z}_q}\left(\frac{8\omega_{\vec{q}}\sqrt{\omega_{\vec{q}}}~\mathbf a^{(-)}_{\vec{q}}}{1+z_q\bar{z}_q}\right)\\
j^{-}_{\bar{w}}(\tau,w,\bar{w})&=\frac{-L}{(2\pi)^3}\int d\omega_{\vec{q}}~e^{-i\omega_{\vec{q}}L\left(\tau-\pi/2\right)}\int d^2z_q\frac{1}{(\bar{z}_q-\bar{w})}\partial_{\bar{z}_q}\left(\frac{8\omega_{\vec{q}}\sqrt{\omega_{\vec{q}}}~\mathbf a^{(-)}_{\vec{q}}}{1+z_q\bar{z}_q}\right)\\
\partial_{\bar{z}_q}j^{-}_{{z}_q}(\tau,z_q,\bar{z}_q)&=\frac{L}{(2\pi)^2}\int d\omega_{\vec{q}}~e^{-i\omega_{\vec{q}}L\left(\tau-\pi/2\right)}\partial_{{z}_q}\left(\frac{8\omega_{\vec{q}}\sqrt{\omega_{\vec{q}}}~\mathbf a^{(+)}_{\vec{q}}}{1+z_q\bar{z}_q}\right)\\
j^{-}_{{w}}(\tau,w,\bar{w})&=\frac{-L}{(2\pi)^3}\int d\omega_{\vec{q}}~e^{-i\omega_{\vec{q}}L\left(\tau-\pi/2\right)}\int d^2z_q\frac{1}{({z}_q-{w})}\partial_{{z}_q}\left(\frac{8\omega_{\vec{q}}\sqrt{\omega_{\vec{q}}}~\mathbf a^{(+)}_{\vec{q}}}{1+z_q\bar{z}_q}\right)
\end{split}
\end{equation}
Using these expressions we can write \eqref{j_tau_a} in terms of creation/annihilation operator of photon. Then using the expression of \eqref{exp_I_j} and putting that into \eqref{a_dagger_f} we will have the dressed creation operator in terms of creation/annihilation operator of scalar field and photon. The final expression is a bit long to be presented here. We have written it in \ref{appendixA}.
\section{Conclusions and future directions}\label{conclusion}
\label{Conclusions}
We conclude by summarizing our findings and listing some future directions. In this paper, we have constructed a version of the Faddeev-Kulish dressed state in terms of CFT operators. This has been done by utilizing AdS dressed fields and existing maps between the photon creation operators and the CFT current operator.  
\subsection*{Future directions:}

There are several future directions related to this work.
\subsection*{Construction of BMS supertranslation charge in the flat limit of AdS/CFT:} 
The Faddeev-Kulish dressing can be realized in terms of the eigenstate of BMS supertranslation charge \cite{Choi:2017ylo}. It would be interesting to construct the BMS supertranslation charge in the flat limit of AdS/CFT. This would give an alternative way to arrive the Faddeev-Kulish dressing from AdS/CFT. 

\subsection*{Correction to the Faddeev-Kulish dressed state:}
It would be interesting to analyze the $\mathcal{O}\Big(\frac{1}{L^2}\Big)$ correction to the Faddeev-Kulish dressed state. Recently, in a paper \cite{Banerjee:2022oll} $\mathcal{O}\Big(\frac{1}{L^2}\Big)$ correction to the soft photon theorem was studied where they derive the $\mathcal{O}\Big(\frac{1}{L^2}\Big)$ corrections of the gauge field HKLL kernels. We can use this to calculate the $\mathcal{O}\Big(\frac{1}{L^2}\Big)$ to the Faddeev-Kulish dressed state. We can further study the $\mathcal{O}\Big(\frac{1}{L^2}\Big)$ correction to the free massive scalar field mode as well. We hope to address this question in an upcoming work\cite{future2}.      

\subsection*{CCFT from averaging over global time of AdS/CFT:}
Using the mapping between the creation/annihilation operators and CFT operators \cite{Hijano:2020szl}, we can construct $\mathcal{S}$-matrix from CFT correlators. Now, we can further perform a Mellin transform/convolute with the bulk-to-boundary propagators to get the celestial amplitude in CCFT. It would be exciting to get the CCFT correlators from flat limit of $CFT_3$ correlators\cite{future3}. Also, we can study $\mathcal{O}\Big(\frac{1}{L}\Big)$ correction to the $\mathcal{S}$-matrix from $CFT_3$ correlators and relate it to `\textit{\textbf{moving away from CCFT}}'\cite{future3}. From this, we can translate the IR divergence in $CFT_3$ into CCFT. 

\subsection*{Celestial Integrability and moving away from it:}
For $2d$ $\mathcal{S}$-matrix, integrability can be formulated in terms of the Yang-Baxter equation which expresses $3 \to 3$ scattering processes in terms of $2 \to 2$. Recently, celestial amplitudes are explored for integrable theories\cite{Duary:2022onm,Kapec:2022xjw}. It's interesting to find the map between scattering states and CFT operators in the flat limit of $AdS_2/CFT_1$. We can also try to see the $\mathcal{O}\Big(\frac{1}{L}\Big)$ correction to the flat space $\mathcal{S}$-matrix and CFT correlators. After translating into celestial space, we can try to see the Yang-Baxter equation in celestial space and try to see the corrections to it and how far we can go to solve the $\mathcal{S}$-matrix exactly\cite{future4}.  

\section*{Acknowledgements}
We would like to thank Shamik Banerjee, Sayantani Bhattacharyya, Sangmin Choi, Hofie Sigridar Hannesdottir, Daniel Kabat, Chethan Krishnan, Alok Laddha, Gilad Lifschytz, R.Loganayagam, Pronobesh Maity, Priyadarshi Paul, Suvrat Raju, Pabitra Ray, Iain Stewart, Ashoke Sen, Becher Thomas for useful discussions. S.D. would especially like to thank Alok Laddha and R.Loganayagam for giving him mental support throughout this work. M.P would especially like to thank Sayantani Bhattacharyya and Alok Laddha for encouragement throughout the course of the work. The work of S.D. is supported by the Department of Atomic Energy, Government of India, under project no. RTI4001. 
\appendix

\section{Detailed derivation of some equations in section \ref{dressedaa}}\label{appendixA}
We will use complex coordinates on the two sphere, where the standard coordinates $\theta\in \left[0,\pi\right]$ and $\phi\in (0,2\pi]$ is related to the complex coordinates $z$ and $\bar{z}$ as follows.
\begin{equation}
\begin{split}
\cos\theta=\frac{1-z\bar{z}}{1+z\bar{z}},~~\sin\theta \cos\phi=\frac{z+\bar{z}}{1+z\bar{z}},~~\sin\theta \sin\phi=-i\frac{z-\bar{z}}{1+z\bar{z}}~.
\end{split}
\end{equation}
Then the metric in these coordinates reads
\begin{equation}
\begin{split}
ds^2=\frac{4dz d\bar{z}}{\left(1+z\bar{z}\right)^2}.
\end{split}
\end{equation}
 We have the photon creation operator from \citep{Hijano:2020szl} as
\begin{equation}\label{aaja}
\begin{split}
&\sqrt{2\omega_{\vec{q}}}~\mathbf a^{\dagger (-)}_{\vec{q}}=\frac{-1}{4\omega_{\vec{q}}}\frac{1+z_q\bar{z}_q}{\sqrt{2}}\int d\tau^{\prime}~e^{i\omega_{\vec{q}}L\left(\frac{\pi}{2}-\tau^{\prime}\right)}\int d^2z^{\prime}\frac{1}{\left(z_q-z^{\prime}\right)^2}j^+_{\bar{z}^{\prime}}(\tau^{\prime},z^{\prime},\bar{z}^{\prime})\\
&\sqrt{2\omega_{\vec{q}}}~ \mathbf a^{\dagger (+)}_{\vec{q}}=\frac{-1}{4\omega_{\vec{q}}}\frac{1+z_q\bar{z}_q}{\sqrt{2}}\int d\tau^{\prime}~e^{i\omega_{\vec{q}}L\left(\frac{\pi}{2}-\tau^{\prime}\right)}\int d^2z^{\prime}\frac{1}{\left(\bar{z}_q-\bar{z}^{\prime}\right)^2}j^+_{{z}^{\prime}}(\tau^{\prime},z^{\prime},\bar{z}^{\prime})\\
\end{split}
\end{equation}
Note that here the $+$ symbol in $j^+_{\bar{z}^{\prime}}$ stands for the positive frequency mode of the gauge field. We first invert this formula to write $j^+_{\bar{z}^{\prime}}$ in terms of the creation operator of photon. Acting with a $\partial_{\bar{z}_q}$ on both sides of the first equation of \eqref{aaja} we have
\begin{equation}
\begin{split}
\partial_{\bar{z}_q}\left(\frac{(-8)\omega_{\vec{q}}\sqrt{\omega_{\vec{q}}}~\mathbf a^{\dagger (-)}_{\vec{q}}}{1+z_q\bar{z}_q}\right)=\int d\tau^{\prime}~e^{i\omega_{\vec{q}}L\left(\pi/2-\tau^{\prime}\right)}\int d^2z^{\prime}~\partial_{\bar{z}_q}\frac{1}{(z_q-z^{\prime})^2}~j^{+}_{\bar{z}^{\prime}}\left(\tau^{\prime},z^{\prime},\bar{z}^{\prime}\right)
\end{split}
\end{equation}
Now we use the following identity
\begin{equation}
\begin{split}
\partial_{\bar{z}}\frac{1}{(z-z_q)^{n+1}}=(2\pi)\frac{(-1)^{n}}{n!}\partial_z^n\delta^{(2)}(z,z_q)
\end{split}
\end{equation}
Then we have
\begin{equation}
\begin{split}
\partial_{z_q}j^{+}_{\bar{z}_q}(\tau,z_q,\bar{z}_q)=\frac{L}{(2\pi)^2}\int d\omega_{\vec{q}}~e^{i\omega_{\vec{q}}L\left(\tau-\pi/2\right)}\partial_{\bar{z}_q}\left(\frac{8\omega_{\vec{q}}\sqrt{\omega_{\vec{q}}}~\mathbf a^{\dagger (-)}_{\vec{q}}}{1+z_q\bar{z}_q}\right)
\end{split}
\end{equation}
Now we multiply both sides of this equation by $\frac{1}{\bar{z}_q-\bar{w}}$ and then integrate with respect to $z_q$, then we have
\begin{equation}
\begin{split}
j^{+}_{\bar{w}}(\tau,w,\bar{w})=\frac{-L}{(2\pi)^3}\int d\omega_{\vec{q}}~e^{i\omega_{\vec{q}}L\left(\tau-\pi/2\right)}\int d^2z_q\frac{1}{(\bar{z}_q-\bar{w})}\partial_{\bar{z}_q}\left(\frac{8\omega_{\vec{q}}\sqrt{\omega_{\vec{q}}}~\mathbf a^{\dagger (-)}_{\vec{q}}}{1+z_q\bar{z}_q}\right)
\end{split}
\end{equation}
where we have used the following identity
\begin{equation}
\begin{split}
\partial_z\frac{1}{(\bar{z}-\bar{z}_q)}=(2\pi)\delta^{(2)}(z_q,z).
\end{split}
\end{equation}
Following similar steps we have
\begin{equation}
\begin{split}
\partial_{\bar{z}_q}j^{+}_{{z}_q}(\tau,z_q,\bar{z}_q)&=\frac{L}{(2\pi)^2}\int d\omega_{\vec{q}}~e^{i\omega_{\vec{q}}L\left(\tau-\pi/2\right)}\partial_{{z}_q}\left(\frac{8\omega_{\vec{q}}\sqrt{\omega_{\vec{q}}}~\mathbf a^{\dagger (+)}_{\vec{q}}}{1+z_q\bar{z}_q}\right)\\
j^{+}_{{w}}(\tau,w,\bar{w})&=\frac{-L}{(2\pi)^3}\int d\omega_{\vec{q}}~e^{i\omega_{\vec{q}}L\left(\tau-\pi/2\right)}\int d^2z_q\frac{1}{({z}_q-{w})}\partial_{{z}_q}\left(\frac{8\omega_{\vec{q}}\sqrt{\omega_{\vec{q}}}~\mathbf a^{\dagger (+)}_{\vec{q}}}{1+z_q\bar{z}_q}\right)
\end{split}
\end{equation}
Similarly, we have the photon annihilation operator from \citep{Hijano:2020szl} as
\begin{equation}\label{aajaa}
\begin{split}
&\sqrt{2\omega_{\vec{q}}}~\mathbf a^{ (-)}_{\vec{q}}=\frac{-1}{4\omega_{\vec{q}}}\frac{1+z_q\bar{z}_q}{\sqrt{2}}\int d\tau^{\prime}~e^{-i\omega_{\vec{q}}L\left(\frac{\pi}{2}-\tau^{\prime}\right)}\int d^2z^{\prime}\frac{1}{\left(z^{\prime}-z_q\right)}\partial_{z^{\prime}}j^-_{\bar{z}^{\prime}}(\tau^{\prime},z^{\prime},\bar{z}^{\prime})\\
&\sqrt{2\omega_{\vec{q}}}~ \mathbf a^{(+)}_{\vec{q}}=\frac{-1}{4\omega_{\vec{q}}}\frac{1+z_q\bar{z}_q}{\sqrt{2}}\int d\tau^{\prime}~e^{-i\omega_{\vec{q}}L\left(\frac{\pi}{2}-\tau^{\prime}\right)}\int d^2z^{\prime}\frac{1}{\left(\bar{z}^{\prime}-\bar{z}_q\right)}\partial_{\bar{z}^{\prime}}j^-_{{z}^{\prime}}(\tau^{\prime},z^{\prime},\bar{z}^{\prime})\\
\end{split}
\end{equation}
Inverting these equations following similar steps as before we have
\begin{equation}
\begin{split}
\partial_{{z}_q}j^{-}_{\bar{z}_q}(\tau,z_q,\bar{z}_q)&=\frac{L}{(2\pi)^2}\int d\omega_{\vec{q}}~e^{-i\omega_{\vec{q}}L\left(\tau-\pi/2\right)}\partial_{\bar{z}_q}\left(\frac{8\omega_{\vec{q}}\sqrt{\omega_{\vec{q}}}~\mathbf a^{(-)}_{\vec{q}}}{1+z_q\bar{z}_q}\right)\\
j^{-}_{\bar{w}}(\tau,w,\bar{w})&=\frac{-L}{(2\pi)^3}\int d\omega_{\vec{q}}~e^{-i\omega_{\vec{q}}L\left(\tau-\pi/2\right)}\int d^2z_q\frac{1}{(\bar{z}_q-\bar{w})}\partial_{\bar{z}_q}\left(\frac{8\omega_{\vec{q}}\sqrt{\omega_{\vec{q}}}~\mathbf a^{(-)}_{\vec{q}}}{1+z_q\bar{z}_q}\right)\\
\partial_{\bar{z}_q}j^{-}_{{z}_q}(\tau,z_q,\bar{z}_q)&=\frac{L}{(2\pi)^2}\int d\omega_{\vec{q}}~e^{-i\omega_{\vec{q}}L\left(\tau-\pi/2\right)}\partial_{{z}_q}\left(\frac{8\omega_{\vec{q}}\sqrt{\omega_{\vec{q}}}~\mathbf a^{(+)}_{\vec{q}}}{1+z_q\bar{z}_q}\right)\\
j^{-}_{{w}}(\tau,w,\bar{w})&=\frac{-L}{(2\pi)^3}\int d\omega_{\vec{q}}~e^{-i\omega_{\vec{q}}L\left(\tau-\pi/2\right)}\int d^2z_q\frac{1}{({z}_q-{w})}\partial_{{z}_q}\left(\frac{8\omega_{\vec{q}}\sqrt{\omega_{\vec{q}}}~\mathbf a^{(+)}_{\vec{q}}}{1+z_q\bar{z}_q}\right)
\end{split}
\end{equation}
The expression for the $\tau$ component of the boundary current for positive and negative frequency modes are
\begin{equation}\label{j_tau_a}
\begin{split}
j_{\tau^{\prime}}^{\pm}=\int_{0}^{\tau^{\prime}}d{\tilde\tau}\Bigg(\frac{1}{2}(1+z\bar{z})^2(\partial_{z}j_{\bar{z}}^{\pm}+\partial_{\bar{z}}j_{z}^{\pm})+(1+z\bar{z})\left(zj_{z}^{\pm}+\bar{z}j_{\bar{z}}^{\pm}\right)\Bigg)~.
\end{split}
\end{equation}
Putting the expressions for $\partial_{z}j_{\bar{z}}^{\pm},  \partial_{\bar{z}}j_{z}^{\pm}, j_{z}^{\pm}$ and $j_{\bar{z}}^{\pm}$ we have
\begin{equation}
\small 
\begin{split}
j_{\tau}^{+}
&=\frac{1}{2}(1+z^{\prime}\bar{z}^{\prime})^2\Bigg[\frac{L}{(2\pi)^2}\int d\omega_{\vec{q}}~ \frac{1}{iL\omega_{\vec{q}}}e^{-\frac{1}{2} i\pi L \omega_{\vec{q}}}(e^{iL\tau \omega_{\vec{q}}}-1) \partial_{\bar{z}_q}~\Bigg(\sqrt{2 \omega_{\vec{q}}}~ 4\omega_{\vec{q}} ~\mathbf a_{\vec{q}}^{\dagger(-)}\frac{\sqrt{2}}{1+z_{q}\bar{z}_{q}}
\Bigg)\Bigg|_{z_q=z^{\prime}}\\
&~~~~~~~~~~~~~~~~~~~~~+\frac{L}{(2\pi)^2}\int d\omega_{\vec{q}}~ \frac{1}{iL\omega_{\vec{q}}}e^{-\frac{1}{2} i\pi L \omega_{\vec{q}}}(e^{iL\tau \omega_{\vec{q}}}-1) \partial_{z_q}~\Bigg(\sqrt{2 \omega_{\vec{q}}}~ 4\omega_{\vec{q}} ~\mathbf a_{\vec{q}}^{\dagger(+)}\frac{\sqrt{2}}{1+z_{q}\bar{z}_{q}}\Bigg)\Bigg|_{z_q=z^{\prime}}
\Bigg]\\
&+\bar{z}^{\prime}(1+z^{\prime}\bar{z}^{\prime})\Bigg[
-\frac{L}{(2\pi)^3}\int d\omega_{\vec{q}}~\frac{1}{iL\omega_{\vec{q}}}e^{-\frac{1}{2} i\pi L \omega_{\vec{q}}}(e^{iL\tau \omega_{\vec{q}}}-1)\int d^2 z_q\frac{1}{\bar{z}_q-\bar{z}^{\prime}} \partial_{\bar{z}_q}~\Bigg(\sqrt{2 \omega_{\vec{q}}}~ 4\omega_{\vec{q}} ~\mathbf a_{\vec{q}}^{\dagger(-)}\frac{\sqrt{2}}{1+z_{q}\bar{z}_{q}}\Bigg)
\Bigg]\\
&+z^{\prime}(1+z^{\prime}\bar{z}^{\prime})\Bigg[-\frac{L}{(2\pi)^3}\int d\omega_{\vec{q}}~ \frac{1}{iL\omega_{\vec{q}}}e^{-\frac{1}{2} i\pi L \omega_{\vec{q}}}(e^{iL\tau \omega_{\vec{q}}}-1)\int d^2 z_q \frac{1}{z_q-z^{\prime}}  \partial_{z_q}~\Bigg(\sqrt{2 \omega_{\vec{q}}}~ 4\omega_{\vec{q}} ~\mathbf a_{\vec{q}}^{\dagger(+)}\frac{\sqrt{2}}{1+z_{q}\bar{z}_{q}}\Bigg)~\Bigg]~~~,
\end{split}
\end{equation}

\begin{equation}
\small 
\begin{split}
j_{\tau}^{-}
&=\frac{1}{2}(1+z^{\prime}\bar{z}^{\prime})^2\Bigg[\frac{L}{(2\pi)^2}\int d\omega_{\vec{q}}~ \frac{1}{-iL\omega_{\vec{q}}}e^{\frac{1}{2} i\pi L \omega_{\vec{q}}}(e^{-iL\tau \omega_{\vec{q}}}-1) \partial_{\bar{z}_q}~\Bigg(\sqrt{2 \omega_{\vec{q}}}~ 4\omega_{\vec{q}} ~\mathbf a_{\vec{q}}^{(-)}\frac{\sqrt{2}}{1+z_{q}\bar{z}_{q}}
\Bigg)\Bigg|_{z_q=z^{\prime}}\\
&~~~~~~~~~~~~~~~~~~~~~+\frac{L}{(2\pi)^2}\int d\omega_{\vec{q}}~ \frac{1}{-iL\omega_{\vec{q}}}e^{\frac{1}{2} i\pi L \omega_{\vec{q}}}(e^{-iL\tau \omega_{\vec{q}}}-1) \partial_{z_q}~\Bigg(\sqrt{2 \omega_{\vec{q}}}~ 4\omega_{\vec{q}} ~\mathbf a_{\vec{q}}^{(+)}\frac{\sqrt{2}}{1+z_{q}\bar{z}_{q}}\Bigg)\Bigg|_{z_q=z^{\prime}}
\Bigg]\\
&+\bar{z}^{\prime}(1+z^{\prime}\bar{z}^{\prime})\Bigg[-
\frac{L}{(2\pi)^3}\int d\omega_{\vec{q}}~\frac{1}{-iL\omega_{\vec{q}}}e^{\frac{1}{2} i\pi L \omega_{\vec{q}}}(e^{-iL\tau \omega_{\vec{q}}}-1)\int d^2 z_q\frac{1}{\bar{z}_q-\bar{z}^{\prime}} \partial_{\bar{z}_q}~\Bigg(\sqrt{2 \omega_{\vec{q}}}~ 4\omega_{\vec{q}} ~\mathbf a_{\vec{q}}^{(-)}\frac{\sqrt{2}}{1+z_{q}\bar{z}_{q}}\Bigg)
\Bigg]\\
&+z^{\prime}(1+z^{\prime}\bar{z}^{\prime})\Bigg[-\frac{L}{(2\pi)^3}\int d\omega_{\vec{q}}~ \frac{1}{-iL\omega_{\vec{q}}}e^{\frac{1}{2} i\pi L \omega_{\vec{q}}}(e^{-iL\tau \omega_{\vec{q}}}-1)\int d^2 z_q \frac{1}{z_q-z^{\prime}}  \partial_{z_q}~\Bigg(\sqrt{2 \omega_{\vec{q}}}~ 4\omega_{\vec{q}} ~\mathbf a_{\vec{q}}^{(+)}\frac{\sqrt{2}}{1+z_{q}\bar{z}_{q}}\Bigg)~\Bigg]~~~.
\end{split}
\end{equation}
Putting these expressions in \eqref{a_dagger_f} and after some simplification, at $\mathcal{O}(q)$ we have the expression for the dressed state as
\begin{equation}
\small 
\begin{split}
\sqrt{2\omega_{\vec{p}}}~\widetilde{a}^{\dagger}_{\omega_{\vec{p}}}
\left |0\right>&=\frac{\tilde{{C}}}{{C}} ~~\Bigg[-iq \Bigg\{\frac{L}{2}(1+z^{\prime}\bar{z}^{\prime})^2 \times \\
& \Bigg[\int d\omega_{\vec{q}}~ \Bigg\{ \frac{1}{iL \omega_{\vec{q}}}\Big(e^{\frac{L}{2}\Big[-i\pi \omega_{\vec{q}}-\omega_{\vec{p}}\log\Big(\frac{\omega_{\vec{p}}+m}{\omega_{\vec{p}}-m}\Big)+(\omega_{\vec{p}}+\omega_{\vec{q}})\log\Big(\frac{\omega_{\vec{p}}+\omega_{\vec{q}}+m}{\omega_{\vec{p}}+\omega_{\vec{q}}-m}\Big)\Big]}\sqrt{2(\omega_{\vec{p}}+\omega_{\vec{q}})}a^{\dagger}_{\omega_{\vec{p}}+\omega_{\vec{q}}}-\sqrt{2\omega_{\vec{p}}}a^{\dagger}_{\omega_{\vec{p}}}\Big)\\
&+ie^{i\omega_{\vec{p}}L\Big[\frac{\pi}{2}+\frac{i}{2}\log\Big(\frac{\omega_{\vec{p}}+m}{\omega_{\vec{p}}-m}\Big)\Big]}\frac{\partial}{\partial \Delta_{\vec{p}}}\Big(\sqrt{2\Delta_{\vec{p}}}a^{\dagger}_{\Delta_{\vec{p}}}e^{-i \Delta_{\vec{p}}L\Big[ \frac{\pi}{2}+\frac{i}{2}\log\Big(\frac{\Delta_{\vec{p}}+m}{\Delta_{\vec{p}}-m}\Big)\Big]}\Big)\Bigg|_{\Delta_{\vec{p}}=\omega_{\vec{p}}}\Bigg\}\\ 
&~~~~~~~~~\times -\frac{1}{(2\pi)^2} \frac{1}{iL\omega_{\vec{q}}}  e^{-\frac{1}{2} i\pi L \omega_{\vec{q}}}\partial_{\bar{z}_q}~\Bigg(\sqrt{2 \omega_{\vec{q}}}~ 4\omega_{\vec{q}} ~\mathbf a_{\vec{q}}^{\dagger(-)}\frac{\sqrt{2}}{1+z_{q}\bar{z}_{q}}
\Bigg)\Bigg|_{z_q=z^{\prime}}\\
&-\int d\omega_{\vec{q}}~\Bigg\{ \frac{1}{iL \omega_{\vec{q}}}\Big(e^{\frac{L}{2}\Big[-i\pi \omega_{\vec{q}}-\omega_{\vec{p}}\log\Big(\frac{\omega_{\vec{p}}+m}{\omega_{\vec{p}}-m}\Big)+(\omega_{\vec{p}}+\omega_{\vec{q}})\log\Big(\frac{\omega_{\vec{p}}+\omega_{\vec{q}}+m}{\omega_{\vec{p}}+\omega_{\vec{q}}-m}\Big)\Big]}\sqrt{2(\omega_{\vec{p}}+\omega_{\vec{q}})}a^{\dagger}_{\omega_{\vec{p}}+\omega_{\vec{q}}}-\sqrt{2\omega_{\vec{p}}}a^{\dagger}_{\omega_{\vec{p}}}\Big)\\
&+ie^{i\omega_{\vec{p}}L\Big[\frac{\pi}{2}+\frac{i}{2}\log\Big(\frac{\omega_{\vec{p}}+m}{\omega_{\vec{p}}-m}\Big)\Big]}\frac{\partial}{\partial \Delta_{\vec{p}}}\Big(\sqrt{2\Delta_{\vec{p}}}a^{\dagger}_{\Delta_{\vec{p}}}e^{-i \Delta_{\vec{p}}L\Big[ \frac{\pi}{2}+\frac{i}{2}\log\Big(\frac{\Delta_{\vec{p}}+m}{\Delta_{\vec{p}}-m}\Big)\Big]}\Big)\Bigg|_{\Delta_{\vec{p}}=\omega_{\vec{p}}}\Bigg\}\\
&~~~~~~~~~\times -\frac{1}{(2\pi)^2}\frac{1}{iL\omega_{\vec{q}}}e^{-\frac{1}{2} i\pi L \omega_{\vec{q}}}\partial_{z_q}~\Bigg(\sqrt{2 \omega_{\vec{q}}}~ 4\omega_{\vec{q}} ~\mathbf a_{\vec{q}}^{\dagger(+)}\frac{\sqrt{2}}{1+z_{q}\bar{z}_{q}}\Bigg)\Bigg|_{z_q=z^{\prime}}
\Bigg]\\
&+\bar{z}^{\prime}(1+z^{\prime}\bar{z}^{\prime})L\\
&\Bigg[
\int d\omega_{\vec{q}}~\Bigg\{ \frac{1}{iL \omega_{\vec{q}}}\Big(e^{\frac{L}{2}\Big[-i\pi \omega_{\vec{q}}-\omega_{\vec{p}}\log\Big(\frac{\omega_{\vec{p}}+m}{\omega_{\vec{p}}-m}\Big)+(\omega_{\vec{p}}+\omega_{\vec{q}})\log\Big(\frac{\omega_{\vec{p}}+\omega_{\vec{q}}+m}{\omega_{\vec{p}}+\omega_{\vec{q}}-m}\Big)\Big]}\sqrt{2(\omega_{\vec{p}}+\omega_{\vec{q}})}a^{\dagger}_{\omega_{\vec{p}}+\omega_{\vec{q}}}-\sqrt{2\omega_{\vec{p}}}a^{\dagger}_{\omega_{\vec{p}}}\Big)\\
&+ie^{i\omega_{\vec{p}}L\Big[\frac{\pi}{2}+\frac{i}{2}\log\Big(\frac{\omega_{\vec{p}}+m}{\omega_{\vec{p}}-m}\Big)\Big]}\frac{\partial}{\partial \Delta_{\vec{p}}}\Big(\sqrt{2\Delta_{\vec{p}}}a^{\dagger}_{\Delta_{\vec{p}}}e^{-i \Delta_{\vec{p}}L\Big[ \frac{\pi}{2}+\frac{i}{2}\log\Big(\frac{\Delta_{\vec{p}}+m}{\Delta_{\vec{p}}-m}\Big)\Big]}\Big)\Bigg|_{\Delta_{\vec{p}}=\omega_{\vec{p}}}\Bigg\}\\
&\times \frac{1}{(2\pi)^3}\frac{1}{iL\omega_{\vec{q}}}e^{-\frac{1}{2} i\pi L \omega_{\vec{q}}}\int d^2 z_q\frac{1}{\bar{z}_q-\bar{z}^{\prime}} \partial_{\bar{z}_q}~\Bigg(\sqrt{2 \omega_{\vec{q}}}~ 4\omega_{\vec{q}} ~\mathbf a_{\vec{q}}^{\dagger(-)}\frac{\sqrt{2}}{1+z_{q}\bar{z}_{q}}\Bigg)
\Bigg]\\
&+z^{\prime}(1+z^{\prime}\bar{z}^{\prime})L\\
&\Bigg[\int d\omega_{\vec{q}}~\Bigg\{ \frac{1}{iL \omega_{\vec{q}}}\Big(e^{\frac{L}{2}\Big[-i\pi \omega_{\vec{q}}-\omega_{\vec{p}}\log\Big(\frac{\omega_{\vec{p}}+m}{\omega_{\vec{p}}-m}\Big)+(\omega_{\vec{p}}+\omega_{\vec{q}})\log\Big(\frac{\omega_{\vec{p}}+\omega_{\vec{q}}+m}{\omega_{\vec{p}}+\omega_{\vec{q}}-m}\Big)\Big]}\sqrt{2(\omega_{\vec{p}}+\omega_{\vec{q}})}a^{\dagger}_{\omega_{\vec{p}}+\omega_{\vec{q}}}-\sqrt{2\omega_{\vec{p}}}a^{\dagger}_{\omega_{\vec{p}}}\Big)\\
&+ie^{i\omega_{\vec{p}}L\Big[\frac{\pi}{2}+\frac{i}{2}\log\Big(\frac{\omega_{\vec{p}}+m}{\omega_{\vec{p}}-m}\Big)\Big]}\frac{\partial}{\partial \Delta_{\vec{p}}}\Big(\sqrt{2\Delta_{\vec{p}}}a^{\dagger}_{\Delta_{\vec{p}}}e^{-i \Delta_{\vec{p}}L\Big[ \frac{\pi}{2}+\frac{i}{2}\log\Big(\frac{\Delta_{\vec{p}}+m}{\Delta_{\vec{p}}-m}\Big)\Big]}\Big)\Bigg|_{\Delta_{\vec{p}}=\omega_{\vec{p}}}\Bigg\}\\
&\times\frac{1}{(2\pi)^3}\frac{1}{iL\omega_{\vec{q}}}e^{-\frac{1}{2} i\pi L \omega_{\vec{q}}}\int d^2 z_q \frac{1}{z_q-z^{\prime}}  \partial_{z_q}~\Bigg(\sqrt{2 \omega_{\vec{q}}}~ 4\omega_{\vec{q}} ~\mathbf a_{\vec{q}}^{\dagger(+)}\frac{\sqrt{2}}{1+z_{q}\bar{z}_{q}}\Bigg)~\Bigg]\Bigg\}\Bigg]\left |0\right>~.\\
\end{split}
\end{equation}
\providecommand{\href}[2]{#2}\begingroup\raggedright\endgroup

\end{document}